\documentclass[review]{elsarticle}

\usepackage{lineno,hyperref}
\modulolinenumbers[5]

\usepackage{booktabs}
\usepackage{dcolumn}
\usepackage{hyperref}

\usepackage{amsfonts}
\usepackage{tabularx}
\usepackage{graphicx}
\usepackage{caption}
\usepackage{subcaption}

\journal{}

\begin{document}

\begin{frontmatter}

\title{Multi-column Point-CNN for Sketch Segmentation}

\author[address1]{Fei Wang}
\author[address1]{Shujin Lin}
\author[address2]{Hanhui Li}
\author[address3]{Hefeng Wu}
\author[address1]{Junkun Jiang}
\author[address1]{Ruomei Wang}
\author[address2]{Xiaonan Luo}

\address[address1]{Sun Yat-sen University, Guangzhou, China}
\address[address2]{Guilin University of Electronic Technology, Guilin, China}
\address[address3]{Guangdong University of Foreign Studies, Guangzhou, China}

\begin{abstract}
Traditional sketch segmentation methods mainly rely on handcrafted features and complicate models, and their performance is far from satisfactory due to the abstract representation of sketches. Recent success of Deep Neural Networks (DNNs) in related tasks suggests DNNs could be a practical solution for this problem, yet the suitable datasets for learning and evaluating DNNs are limited. To this end, we introduce SketchSeg, a large dataset consisting of 10,000 pixel-wisely labeled sketches. Besides, due to the lack of colors and textures in sketches, conventional DNNs learned on natural images are not optimal for tackling our problem. Therefore, we further propose the Multi-column Point-CNN (MCPNet), which (1) directly takes sampled points as its input to reduce computational costs, and (2) adopts multiple columns with different filter sizes to better capture the structures of sketches. Extensive experiments validate that the MCPNet is superior to conventional DNNs like FCN. The SketchSeg dataset is publicly available on \url{https://drive.google.com/open?id=1OpCBvkInhxvfAHuVs-spDEppb8iXFC3C}.
\end{abstract}


\end{frontmatter}


\section{Introduction}

\begin{figure}
	\centering
	\includegraphics[width = 0.7\columnwidth]{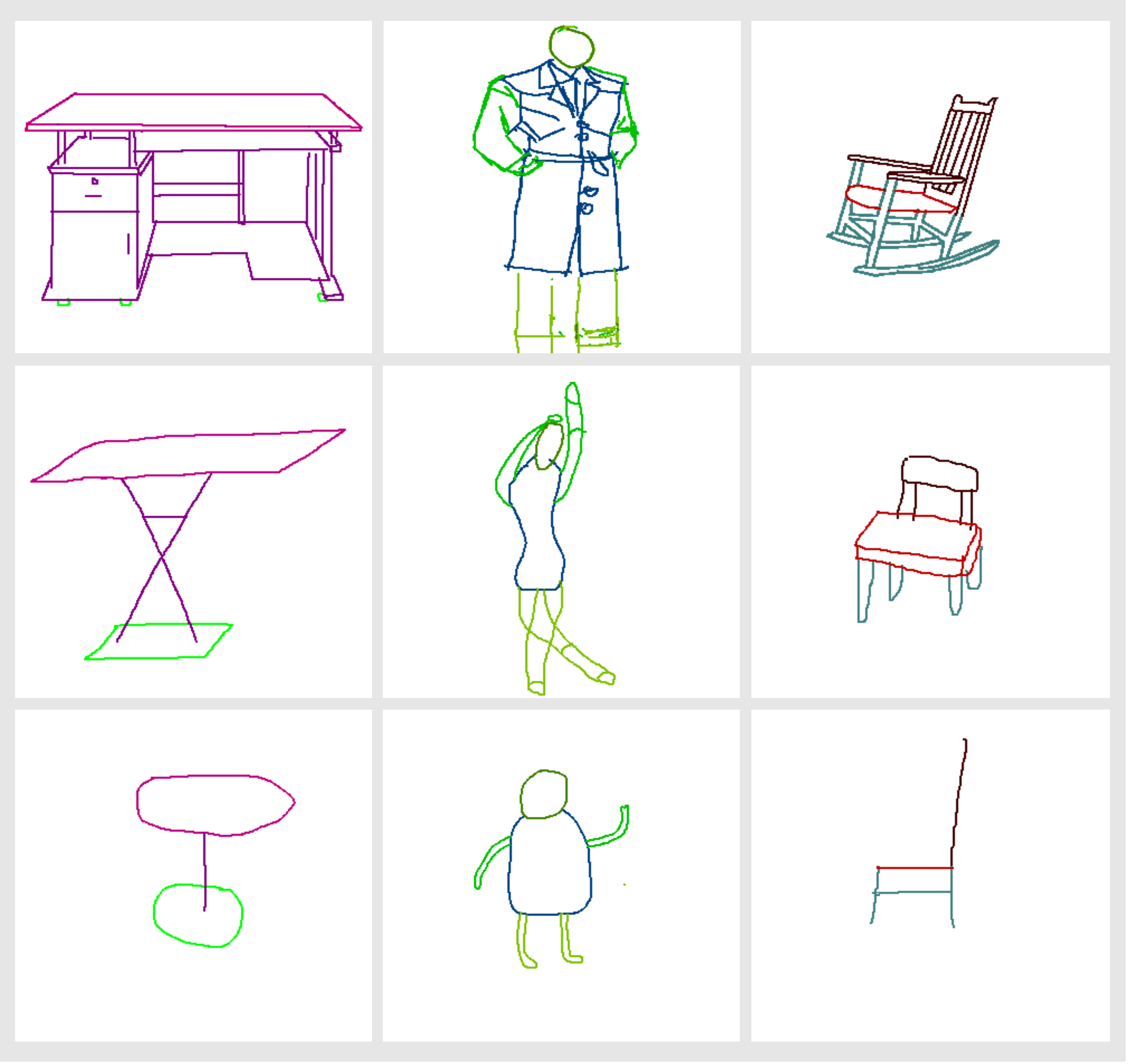}
	\caption{Sketching styles can vary from complex, realistic ones (top) to simple, abstract ones(bottom).}
	\label{fig:differentstyle}
\end{figure}

Free-hand sketches can be considered as a convenient and intuitive tool for human-computer interaction and communication, especially with the widely-used portable devices such as multi-touch tablets and smart phones. Segmenting sketches is an important preprocessing step for many applications, e.g., sketch recognition \cite{yu2015sketch,yu2017sketch}, sketch-based modeling \cite{eitz2011sketch,Eitz2012Sketch,xie2013sketch,wang2017data, shin2007magic,xu2013sketch2scene}, and free-hand sketch synthesis \cite{Li2017Free,ouyang2014cross}. However, the main existing works on sketch segmentation are relied on hand-crafted features and sophisticate models, such as Radial Basis Functions \cite{pu2009automated}, Graph Construction \cite{sun2012free}, Mixed Integer Programming \cite{huang2014data} and Conditional Random Field (CRF) \cite{schneider2016example}. Such methods are not robust enough for modeling individual varieties in artistic skills and styles. For instance, some people tend to draw an object with only its contour, while others may give us the exquisite depictions (Fig. \ref{fig:differentstyle}).

\begin{figure*}[ht]
  \centering
  \includegraphics[width=1\linewidth]{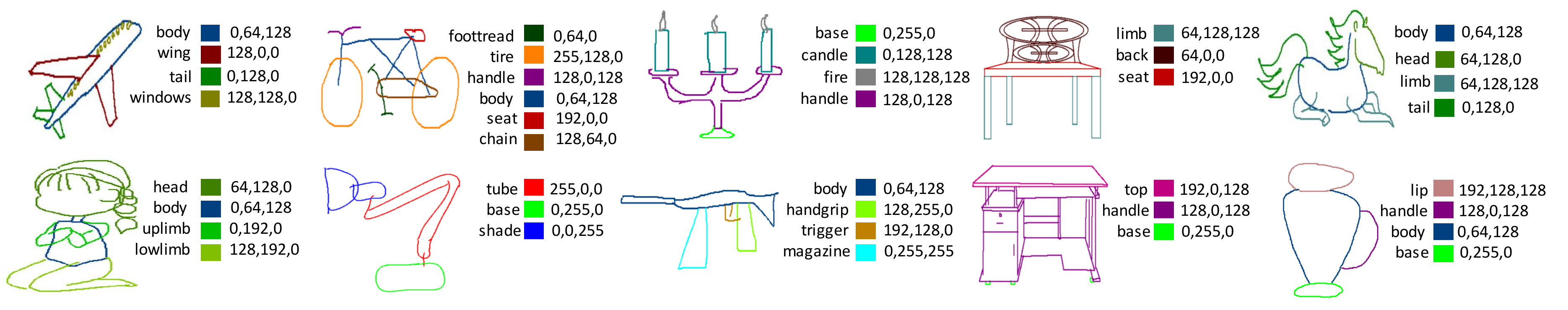}
  \caption{Samples of sketches and their corresponding labels from the SketchSeg database.}\label{fig:sketchexample}
\end{figure*}

Inspired by the recent success of Deep Neural Networks (DNNs) in various semantic segmentation problems \cite{long2015fully,qi2017pointnet,farabet2013learning}, we believe DNNs have the potential to solve the above problem. However, traditional DNNs are more suitable for processing natural images with rich color and texture information, while sketches are just the combination of highly iconic and abstract curves. More importantly, to the best of our knowledge, there is limited sketch datasets available for learning and evaluating DNNs.

To overcome the above challenges, in the paper, we first contribute our sketch segmentation (SketchSeg) database. We follow a previous dataset \cite{huang2014data} to collect sketches belonging to 10 categories, including chair, table, airplane, bicycle, fourleg, lamp, vase, human, candelabrum and rifle. Each category has 1,000 sketches and each sketch is manually labeled with several semantic components. That is, we collect a dataset of 10,000 sketches and their pixel-wise labels, which we believe is large and various enough for perform deep learning. Examples of sketches in our dataset are shown in Fig. \ref{fig:sketchexample}. With the proposed dataset, we conduct extensive experiments to evaluate the performance of conventional DNNs, including U-Net \cite{iglovikov2018ternausnet}, LinkNet \cite{chaurasia2017linknet}, FCN \cite{long2015fully} and PointNet \cite{qi2017pointnet}, so to provide the authentic baselines for sketch segmentation.

Furthermore, we design a Multi-column Point-CNN (MCPNet) that operates directly on the point set sampled from a given sketch and outputs the semantic label of each point. The intention behind our network is that using point set can significantly reduce the computational cost and the side-effect of blank space. Besides, with the proper sampling strategy, we can gather points with similar geometric characteristic together, and maintain the local visual structures of sketches as well. Note that, there might exist sketches with various sizes, and strokes with incomplete adhesion and distortion. Hence the proposed MCPNet adopts multiple columns with filters of different sizes (e.g., large, medium, small,...), so that small filters can better capture the local subtle yet important details, while the large filters prone to model the holistic structure of sketch and be robust to noisy. Consequently combining these filters within the MCPNet makes it a practicable solution for sketch segmentation.

In summary, the contributions of this paper are threefold:
(i) A large new dataset consisting up to 10,000 sketches and their pixel-wise labels is proposed. We believe it will greatly support many interesting researches for sketch analysis.
(ii) We contrive the MCPNet that operates directly on the sampled point set, which can naturally omit the vast blank space in sketches and greatly improve its efficiency.
(iii) The multiple columns in MCPNet adopts filters with different receptive fields, which is able to capture the structural information on various scale levels.

The rest of this paper is organized as follows. In the next section we review the related work. Section \ref{fig:section3} describes the SketchSeg database and our data preprocessing step. Detailed procedure of learning MCPNet is presented in Section \ref{fig:section5}.
Section \ref{fig:section6} demonstrates experimental results. Lastly, we conclude this paper by discussing limitations and possible future applications.

\section{Related work}\label{fig:section2}

In this section, we will first introduce the conventional methods for sketch segmentation. And then we will briefly present the recent progress of utilizing DNNs in relevant tasks, including sketch segmentation, image segmentation and 3D shape analysis.

\subsection{Sketch segmentation}

Segmenting sketches on the semantic level could facilitate applications like sketch-based 3D shape retrieval \cite{xie2013sketch}, scene modeling \cite{xu2013sketch2scene}, part assembly \cite{shen2012structure} and free-hand sketch synthesis \cite{Li2017Free}. And most early study on this topic rely on hand-crafted features with various models from the field of machine learning. For instance, Sun et al. \cite{sun2012free} propose a sketch segmentation framework by combining both the low-level perception and high-level knowledge. Huang et al. \cite{huang2014data} propose a data-driven approach, which uses temporal ordering of strokes as a soft grouping constraint. Schneider et al. \cite{schneider2016example} first estimate the likelihood of each segment belong to a certain component, and then apply the CRF to generate the final global prediction. Li et al. \cite{li2018fast} present a simple and efficient approach based on CNN for semantic segmentation and labeling. They train a DNN to transfer existing segmentations and labeling from 3D models to freehand sketches without requiring numerous annotated sketches.

As to the datasets for sketch segmentation , \cite{wu2018sketchsegnet} and \cite{zou2018sketchyscene} are highly related to our work. What differentiates our dataset from them is our purpose or segmentation level: \cite{wu2018sketchsegnet} focus on the stroke based sketch segmentation, and hence they assume the orders, starting / end points and lengths of all strokes are provided (while sketches are stored as images in our dataset). On the other hand, the aim of \cite{zou2018sketchyscene} is to distinguish sketchy objects in a given scene, namely, their annotations are collected on the object-level. Therefore, in order to locate the semantic components in sketches, it is necessary to build our own dataset.

\subsection{DNNs for image segmentation}

DNNs have been applied into image segmentation successfully, e.g., given a DNN for image classification, one can simply replace its fully connected layers with the fully convolutional ones, and do the fine-tuning to obtain the satisfying performance \cite{long2015fully}. Such Fully Convolutional Network (FCN) can be further improved by incorporating new computational blocks: Zheng et al. \cite{crfasrnn} propose to implement CRF via several basic computational blocks within the framework of DNN; U-Net and its variants \cite{iglovikov2018ternausnet} modify FCN by constructing a strictly symmetric convolution-deconvolution architecture; Chen et al. \cite{chen2018deeplab} introduce the idea of atrous operation, e.g., the astrous convolution and atrous spatial pyramid pooling, to adjust the receptive fields of their DNNs without extra parameters or computations; Mask R-CNN \cite{he2017mask} simply adds a branch for predicting the mask of its target and generates high-quality results. Although the above DNNs have achieved promising results, there is still room for improvement when we considering about utilizing them in sketch segmentation, e.g., the vast blank space in sketches could be omitted.

\subsection{DNNs for 3D shape analysis}
A large variety of DNNs for 3D shape analysis have been developed by the computer vision and graphics community. For instance, Maturana et al. \cite{maturana2015voxnet} and Qi et al. \cite{qi2016volumetric} both consider the volumetric representations of 3D shapes and define their own 3D convolutional filters. Yet these methods was constrained and costly if the data is sparse (e.g., point clouds). Hence, a novel DNN directly taking the coordinates of points as its input, a.k.a. PointNet is proposed \cite{qi2017pointnet}. Even though the accuracy of PointNet is inferior to volumetric based or view based DNNs, its efficiency is notable and has many follow-up methods, such as PointNet++\cite{qi2017pointnet++}, PointNetVLAD\cite{uy2018pointnetvlad} and PointSIFT\cite{jiang2018pointsift}. Our MCPNet derives from PointNet as well, since handling sketches and point clouds have lots in common: one the one hand, we only have the coordinates of the sampled points, neither intensity or texture are available; On the other hand, the sampled points are sparse. Therefore, we propose to tackle our problem within the framework of PointNet. Besides, as we will see in the experimental section, our MCPNet can even outperform traditional DNNs for image segmentation.

\section{The SketchSeg Database}\label{fig:section3}

\begin{table*}[!ht]
\footnotesize
\newcolumntype{Y}{>{\centering\arraybackslash}X}
\centering
\caption{Statistic of number of labeled pixels in SketchSeg, ``C" and ``S" denote component and sketch respectively.}\label{tab:statistic}
\setlength{\tabcolsep}{1mm}{
  \begin{tabular}{p{1.5cm}|p{1cm}p{1cm}p{1.3cm}p{1cm}|p{1cm}p{1cm}p{1.3cm}p{1cm}}
        \toprule
        \hline
 &C-min &C-max&C-mean& C-std&S-min&S-max&S-mean&S-std\\
 \hline
 airplane&39 &3556 & 665.0 & 350.1 & 1146& 7400&2660.2 & 24.7\\
 bicycle  &43 &4334 &497.6 &441.9& 1154 &7967 &2985.7 &28.9\\
 candelabra&48&2791&453.1& 314.1&564&5175&1812.3&26.6\\
 chair&80&5967&739.9 &525.7&536&9235&2219.6&35.1\\
 fourleg&22&2910&704.2&429.8 &1049&5886& 2816.7&26.3 \\
 human&63&3463&659.3&318.6&608&7442& 2637.3&29.9\\
 lamp&62&2155&561.7&222.3&928&3432&1685.2& 19.5\\
 rifle&23&15458&630.7&737.0&487&17600&2522.8 &38.9\\
 table&26&5988&732.5&556.7&528&9114&2197.4&31.2\\
 vase&56&3515&459.8&349.2&703&5085&1839.1&25.4\\
 \hline
    \bottomrule
  \end{tabular}}
\end{table*}

In this section, we introduce the SketchSeg dataset in detail. Our SketchSeg dataset contains sketches of various categories, and each categories has its own semantic components. Specifically, we select the 10 categories defined in \cite{huang2014data}, in order to meet the following criteria: (i) All sketches are from common objects and can be classified clearly, because it is hard for users to draw a ``Godzilla" or tell the difference between pasta and noodle based on sketch. (ii) Each category contains sketches with large intra-diversity, so that we can evaluate the performance of different methods thoroughly.

The 10 categories we selected are chair, table, airplane, bicycle, fourleg, lamp, vase, human, candelabrum and rifle. We then define the semantic components for each categories, which are summarized in Fig. \ref{fig:sketchexample}. All categories have at least 3 semantic components and most of them have 4 components.

With the pre-defined categories and component settings, we recruit 20 volunteers to draw sketches with the Microsoft Paint. Among these volunteers, there are 4 professionals, while the others do not have good painting skills. Every volunteer needs to draw 50 sketches for each category, and to encourage the diversity of their drawings, we do not provide the reference images. The specific drawing settings and procedures are listed as follows:

(i) The volunteer creates a blank canvas of $800 \times 800$ pixels.

(ii) Given a category, we define a unique color (RGB value) for each of its components. Take the ``lamp" class for an example. A lamp can be divided into tube, base and shade, and we denote them by $(255,0,0)$, $(0,255,0)$ and $(0,0,255)$, accordingly.

(iii) The volunteer selects the ``brush tool" with the defined colors to draw the object. The size, position and order of strokes are not restricted, but the volunteer must complete a sketch with all the corresponding components.

\begin{figure}[t]
  \centering
  \includegraphics[width=1\linewidth,height=0.28\linewidth]{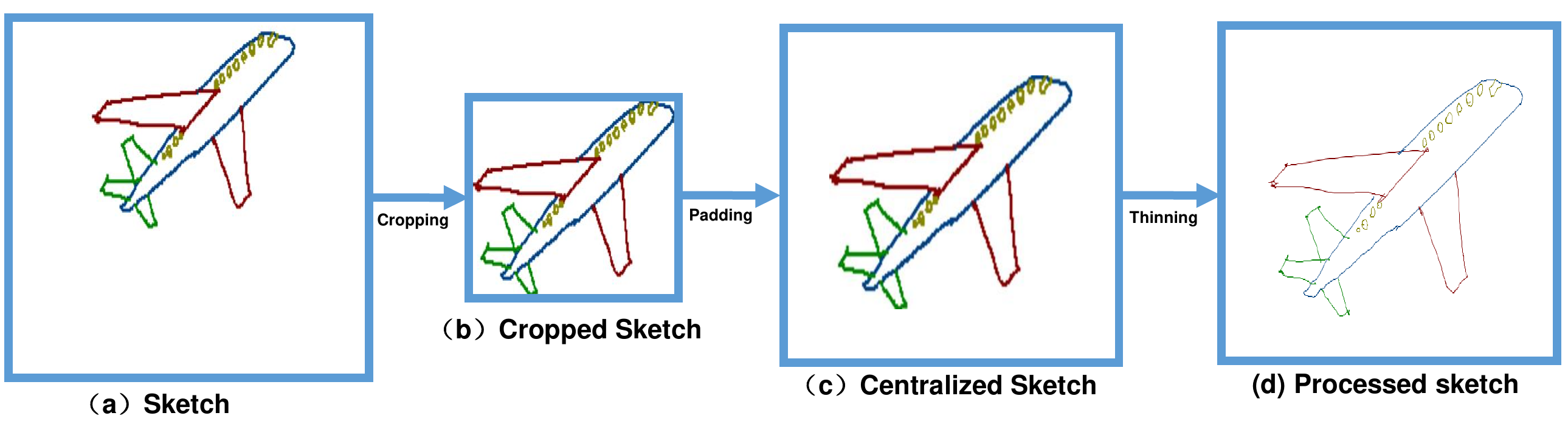}
  \caption{Our work flow of sketch preprocessing.}\label{fig:label}
\end{figure}

(iv) For the convenience of subsequent processing, we crop and pad all sketches to ensure them locate at the center of the canvas. We also perform the morphological thinning operation to ensure the line width of every sketch is 1 pixel, as shown in Figure \ref{fig:label}.

In this way, we have labeled 24 semantic components, and $10 \times 20 \times 50 = 10, 000$ sketches. One interesting fact we have observed from our gathered data is that sketches drawn by professionals are much more realistic than those from ordinary people, for example, the chair and table shown in the fourth column of Fig. \ref{fig:sketchexample}. To quantify the diversity of sketches in our dataset, we propose to calculate the number of labeled pixels with respect to component and sketch. The result is shown in Table \ref{tab:statistic}, which does reflect the arbitrary nature of sketches: for instance, the minimum number of pixel of a sketch in the ``rifle'' class is 487, while the maximum value in the same class can reach 17,600. Such characteristic sketches also makes our dataset become a valuable study target. Our SketchSeg dataset can be accessed on Google Drive, and we hope it could inspire researchers from related communities.

\section{MCPNet Framework}\label{fig:section5}

In this section, we proposed the Multi-Column Point-CNN network (MCPNet) for sketch segmentation. The framework of MCPNet is shown in Fig. \ref{fig:net}: the input sketch is sampled into a 2D point set, which will then be fed into MCPNet to generate the component labels of all sampled points, and thus we obtain the sketch segmentation result. The multi-column architecture is adopted in MCPNet to learn and aggregate multi-scale feature representations. The details of our network will be described in the following subsections.

\subsection{Sketch as arrays}\label{fig:sketcharray}

The freehand sketch is probably the most sparse type of visual image. Its biggest difference from the natural image is that it contains very little image information and is composed of simple curves. Hence utilizing 2D convolutional filters exhaustively on sketch is irrational and unnecessary. Inspired by the work \cite{qi2017pointnet} that utilizes point clouds for 3D model segmentation, we propose to learn feature representation from point sets directly. But our method requires to covert the sketch image into 2D point set rather than 3D point cloud.

\begin{figure}[!h]
  \centering
  \includegraphics[width=0.7\linewidth]{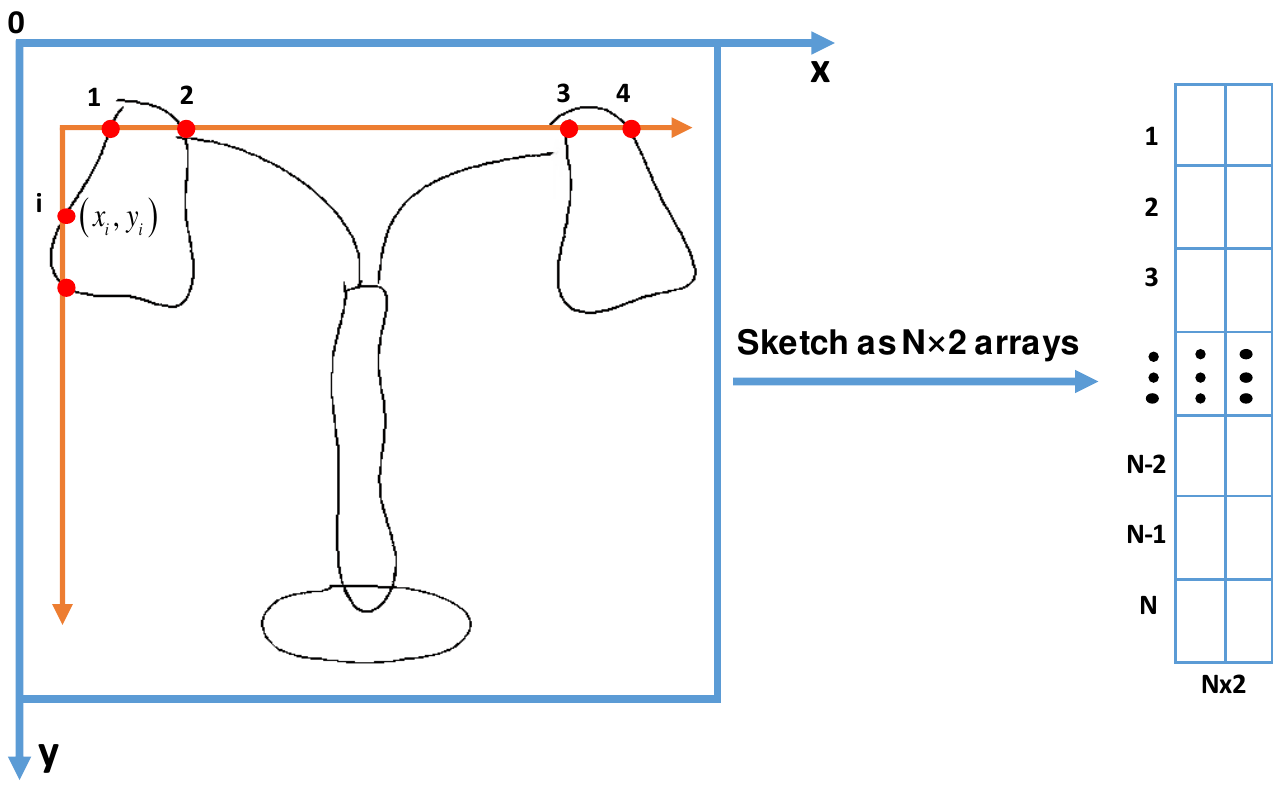}
  \caption{The sketch points are sampled from top-to-bottom and left-to-right.}\label{fig:array1}
\end{figure}

The position and neighborhood relationships of points on the sketch reflect spatial structure information, which are important factors that should not be ignored in sketch segmentation.
For example, in Fig. \ref{fig:array1}, the lamp has a globally symmetrical structure (points 1 and 4, points 2 and 3). And it also has locally symmetrical structures (points 1 and 2, points 3 and 4), which have similar contour structure and relative position information.
Therefore, in this work we collect the coordinates information of sketch by sampling points from left to right and from top to bottom.
Namely, a given sketch is represented as a set of 2D points $P = \{p_{i} \in R^{2}, i = 0, ..., N-1\}$, where $p_{i}$ is a 2D coordinate and $N$ is the number of sampled points. Note that extra visual information like intensity can also be incorporated into $p_{i}$, yet in this paper we only consider 2D coordinates for simplicity.

\subsection{Deep Representation Learning on Point Sets}

Before we introduce the details of our network, we briefly overview the theoretical background of utilizing deep learning on point sets. According to \cite{qi2017pointnet}, let $\mathcal{P} = \mathbb{R}^{N \times 2}$ denote the domain of $P$, and then $\forall \epsilon > 0$, a continuous set function $f:\mathcal{P} \to \mathbb{R}$ can be approximated as follows:
\begin{equation}
 \begin{array}{l}
 |f(P) - \gamma (MAX(h(p_{1}),h(p_{2}),...,h(p_{N})))|<\varepsilon,
  \end{array}
  \label{approx}
\end{equation}
where $\gamma$ and $h$ are both continuous functions, $MAX$ is the element-wise maximum pooling function. Eq.(\ref{approx}) actually indicates that a small perturbation on the order of input points can be neglected by using the $MAX$ function. In our case, it means we can use the basic components of DNNs, such as the convolutional layers to implement $\gamma$ and $h$, and combine them with the max pooling layers to learn our target function $f$.

Specifically, as shown in Fig. \ref{fig:net}, a single column of our MCPNet consists of 3 convolution layers and 1 max pooling layer.
All the 3 convolution layers use kernels of the same size but are with different numbers of channels (i.e., 64, 128 and 1024). After being processed by each of the 3 convolution layers sequentially, the point set $P$ is turned into feature maps $\mathbf{f}_{c1}$ with size of $N\times 64$, $\mathbf{f}_{c2}$ with size of $N\times 128$, and $\mathbf{f}_{c3}$ with size of $N\times 1024$, respectively. Then the feature map $\mathbf{f}_{c3}$ is passed through a max pooling layer to output a feature vector $\mathbf{f}_{g}$ with size of 1024. $\mathbf{f}_{g}$ is supposed to capture the global characteristic of the point set $P$. Then we further combine it with the point-wise information to obtain the feature representation $\mathbf{f}_{P}$ of the point set $P$. To be concrete, we duplicate $\mathbf{f}_{g}$ for $N$ times to form a feature map of size $N\times 1024$, and concatenate it with the feature map $\mathbf{f}_{c1}$  to produce the feature map $\mathbf{f}_{P}$ with size of $N\times 1088$. With the concatenation operation, $\mathbf{f}_{P}$ is able to model not only local structural information, but also global and high-level semantic information.

\subsection{Multi-column Architecture for Sketch Segmentation}

Each sketch has a unique structural form, that is, the combination of points and lines in a two-dimensional plane. However, filters with receptive fields of the same size are unlikely to capture the structural characteristics of sketches. Hence it is more natural to use filters with local receptive fields of different sizes to learn the feature representation from the point set to produce accurate and detailed segmentation.

\begin{figure*}[t]
  \centering
  \includegraphics[width=1\linewidth]{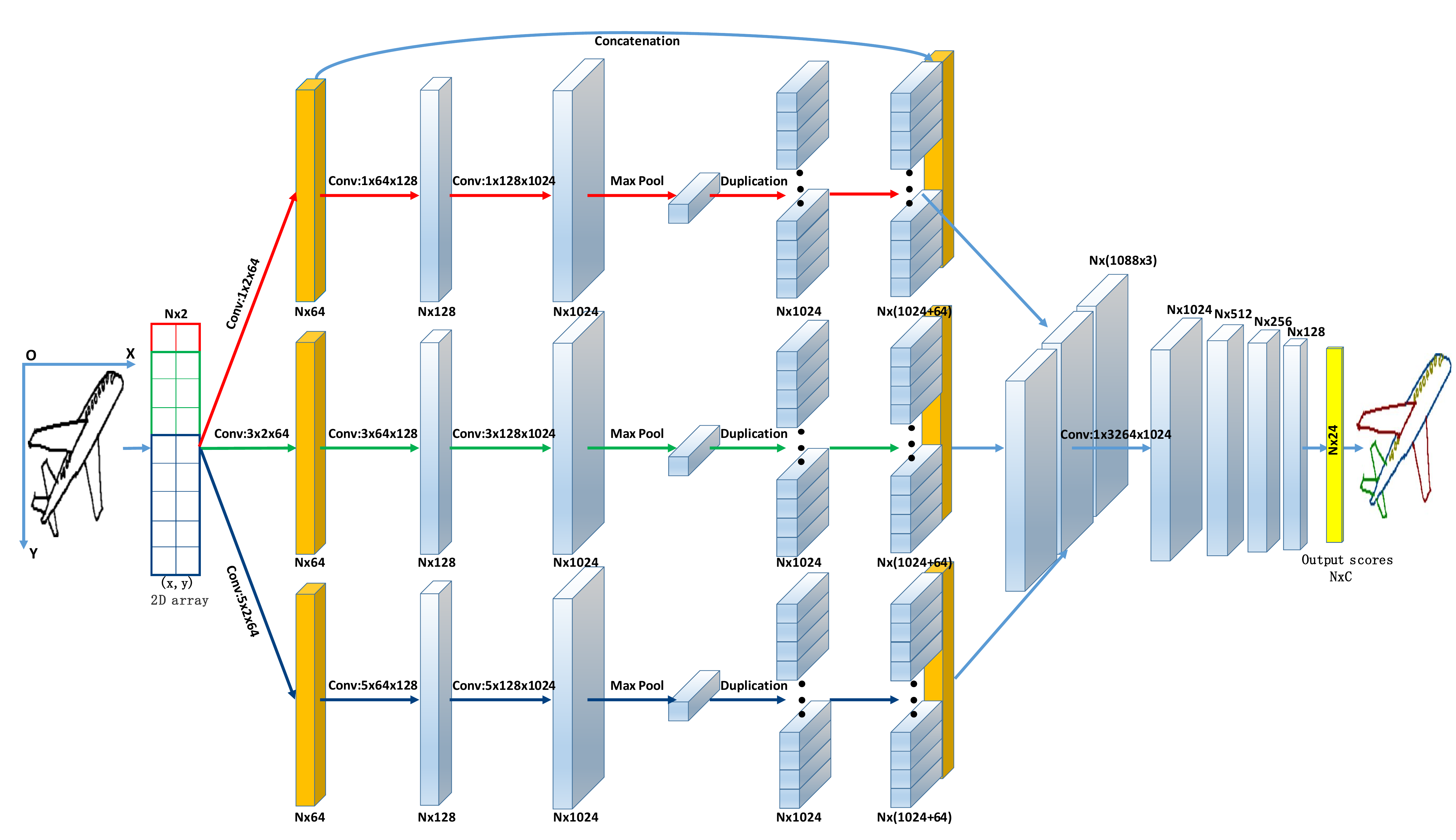}
  \caption{Our network takes coordinates of sketch points as its input, and then aggregates features from multiple columns to generate the segmentation map.}\label{fig:net}
\end{figure*}

Motivated by the success of Multi-column Deep Neural Networks (MDNNs) \cite{cirecsan2012multi}, we incorporate multiple columns into MCPNet, as illustrated in Fig. \ref{fig:net}. Intuitively, filters with large receptive fields are more useful for modeling the symmetric and holistic structure of sketches, while those with small receptive fields are sensitive to the subtle and local information.

Assume there are $K$ columns in our MCPNet. Let the feature map extracted by the $k$-th column be $\mathbf{f}_{P}^{k}$, $k=1,2,\ldots,K$. We concatenate the $K$ feature maps $\{\mathbf{f}_{P}^{k}\}_{k=1}^{K}$ and obtain a feature map $\mathbf{f}_{P}^{a}$ with size of $N\times 1088K$. The feature map $\mathbf{f}_{P}^{a}$ is then passed through several convolutional layers and a softmax layer to generate the final score matrix $S_{P} = (s_{p_{n,c}})\in \mathbb{R}^{N \times C}$, where $C$ is the number of semantic component labels, $s_{p_{n,c}}$ denotes the probability of the $n$-th point belonging to the $c$-th semantic component. The prediction label of each point is determined by the semantic component with the highest probability.

More details of our MCPNet architecture are described below: 3 columns are used in MCPNet, whose filter sizes are $1\times2$, $3\times2$ and $5\times2$, respectively. The five convolutional layers near the output end use kernel of $1 \times 1$, and their channel numbers are 1024, 512, 256, 128 and $C$, respectively. To encourage faster and more stable learning process, batch normalization \cite{ioffe2015batch} with the Rectified Linear Unit (ReLU) is used after all the convolutional layers except the last one in our network.

\subsection{Optimization of MCPNet}

Given a dataset with $M$ training sketch samples $\{X^m, Y^m\}_{m=1}^M$, where $X^m$ is the $m$-th sketch image and $Y^m$ is its corresponding segmentation ground truth, we first turn the training sketch samples into the point set format. That is, the $m$-th training sample is represented as $\{p^m_n,{y}^m_n\}_{n=1}^N$, where $\{p^m_n\}_{n=1}^N$ is the point set sampled from the $m$-th sketch and $\{{y}^m\}_{n=1}^N$ are the corresponding label. $y^m_n\in\{1,...,C\}$ indicates point $p^m_n$ belongs to which of the $C$ components. For point $p^m_n$, we further define a $C$-dimensional one-hot vector $\mathbf{\hat{y}}^m_{n}$, in which the $c$-th element $\hat{y}^m_{n,c}$ is 1 if its label $y^m_n=c$ and all the other elements are 0. Our MCPNet is trained via minimizing the following cross entropy loss function:
\begin{equation}
\mathcal{L} =  - \sum\limits_{m = 1}^M {\sum\limits_{n = 1}^N {\sum\limits_{c = 1}^C {\hat y_{n,c}^m\log s_{n,c}^m}}},
\end{equation}
where ${s_{n,c}^m}$ is the $(n, c)$ element of the score matrix $S_{P}^m$ for the $m$-th training sample. We optimize the above loss $\mathcal{L}$ by the batch-based stochastic gradient descent (SGD) algorithm.

\section{Experiments}\label{fig:section6}

In this section, we will conduct extensive experiments on the proposed SketchSeg dataset. We will compare the proposed MCPNet with several state-of-the-art methods, and provide the detailed ablation study of our network.

\subsection{Setup}

All our experiments are performed on a desktop with a i5 3GHz cpu, 16G RAM and a GTX 1080Ti graphic card. 7,500 sketches from our dataset are used for training while the rest are for testing. We implement 4 cutting-edge methods for comparison, including U-Net \cite{iglovikov2018ternausnet}, LinkNet \cite{chaurasia2017linknet}, FCN \cite{long2015fully} and PointNet \cite{qi2017pointnet}. All these networks are trained with their default architecture and parameter settings. We adopt two metrics \cite{huang2014data} to evaluate the segmentation performance: (i) Pixel-based accuracy (\textbf{P-metric}), which is the fraction of pixels that are assigned with the correct labels; (ii) Component-based accuracy (\textbf{C-metric}), which evaluates the percentage of correctly labeled segments. A component is correctly labeled if 75\% of its pixels are assigned the correct label. Note that for the fair of comparison, all evaluations are performed on the sample points, instead of using the whole sketch images.

Our network is trained for 50 epochs with the SGD algorithm with momentum, where the size of mini-batch, initial learning rate, momentum and weight decay are set to 10, 0.01, 0.9 and 0, respectively. All these parameters are fixed throughout this paper. We use ``MCPNet-x" to denote our network with $x$ columns.

\subsection{Quantitative Analysis}

The segmentation performance evaluated by the P-metric is presented in Table \ref{tab:pixelmetric}. From this result we can see the proposed network can outperform state-of-the-art methods significantly, as it raises the average accuracy from $80.2\%$ to $87.0\%$. Besides, we observe that our performance gain is mostly gained from the ``airplane", ``chair", ``human", ``rifle" and ``table" classes. Considering that the average numbers of labeled pixels of these five classes are larger than those of the rest classes (see Table \ref{tab:statistic}), we can conclude that the proposed network can handle complicate sketches well.

\begin{table*}[!t]
\footnotesize
\newcolumntype{Y}{>{\centering\arraybackslash}X}
  \centering
  \caption{Segmentation performance of MCPNet against state-of-the-art methods with the P-metric.}
  \label{tab:pixelmetric}
  \setlength{\tabcolsep}{1mm}{
  \begin{tabularx}{\textwidth}{Y|YYYY|Y}
        \toprule
        \hline
%
        Method &U-Net&LinkNet&FCN&PointNet&MCPNet-3\\
         \hline
        airplane  &68.9&78.0&78.2&81.0&{\bfseries85.8}\\
        bicycle   &68.1&65.3&71.4&78.0&{\bfseries78.3}\\
        candelabra&89.3&88.3&{\bfseries90.8}&81.1&89.8\\
        chair     &84.0&89.1&86.9&81.0&{\bfseries90.2}\\
        fourleg   &74.1&76.7&80.3&75.5&{\bfseries84.2}\\
        human     &71.9&74.5&75.6&69.2&{\bfseries79.5}\\
        lamp      &92.2&91.2&92.8&86.2&{\bfseries93.2}\\
        rifle     &54.8&59.9&65.2&83.2&{\bfseries87.1}\\
        table     &79.6&82.5&81.4&82.0&{\bfseries92.6}\\
        vase      &89.9&93.8&{\bfseries94.4}&84.8&88.9\\
         \hline
        Average   &77.3&79.9&81.7&80.2&{\bfseries87.0}\\
        \hline
    \bottomrule
  \end{tabularx}}
\end{table*}

The proposed network is superior to conventional methods considering the C-metric as well. As demonstrated in Table \ref{tab:componentmetric}, MCPNet raises the baseline by $1.5\%$. As we know, the C-metric emphasizes on how well a segmentation method could capture the structural information of a sketch, because it requires that the majority of the predicted labels in the same component is correct. However, Taking discrete 2D points directly will lose the structural information to a certain extent, and hence the performance of PointNet is worse than other DNNs. But the multi-column strategy in the proposed network helps us to overcome such disadvantage, and eventually we obtain a performance gain of $9.3\%$, compared with PointNet.

\begin{table*}[!t]
\footnotesize
\newcolumntype{Y}{>{\centering\arraybackslash}X}
  \centering
  \caption{Segmentation performance of MCPNet against state-of-the-art methods with the C-metric.}
  \label{tab:componentmetric}
\setlength{\tabcolsep}{1mm}{
  \begin{tabularx}{\textwidth}{Y|YYYY|Y}
        \toprule
        \hline
%

        Method     &U-Net&LinkNet&FCN&PointNet&MCPNet-3\\
        \hline
         airplane  &52.6&67.7&66.5&67.3&{\bfseries78.8}\\
         bicycle   &49.7&55.7&{\bfseries59.2}&50.9&54.3\\
         candelabra&90.3&89.0&{\bfseries94.5}&67.9&80.5\\
          chair    &81.9&89.2&84.8&77.6&{\bfseries91.1}\\
          fourleg  &54.5&67.2&73.5&60.9&{\bfseries74.8}\\
          human    &62.6&67.9&{\bfseries72.1}&56.6&67.8\\
          lamp     &92.4&92.4&92.5&86.1&{\bfseries95.5}\\
          rifle    &38.9&44.5&54.7&59.7&{\bfseries72.8}\\
          table    &70.1&80.3&75.3&67.5&{\bfseries87.4}\\
          vase     &90.7&96.6&{\bfseries98.1}&78.9&83.1\\
           \hline
          Average  &68.4&75.0&77.1&67.3&{\bfseries78.6}\\
        \hline
    \bottomrule
  \end{tabularx}}
\end{table*}

\begin{figure*}[!ht]
	\centering
	\begin{subfigure}{0.45\textwidth}
		\includegraphics[width = 1\textwidth]{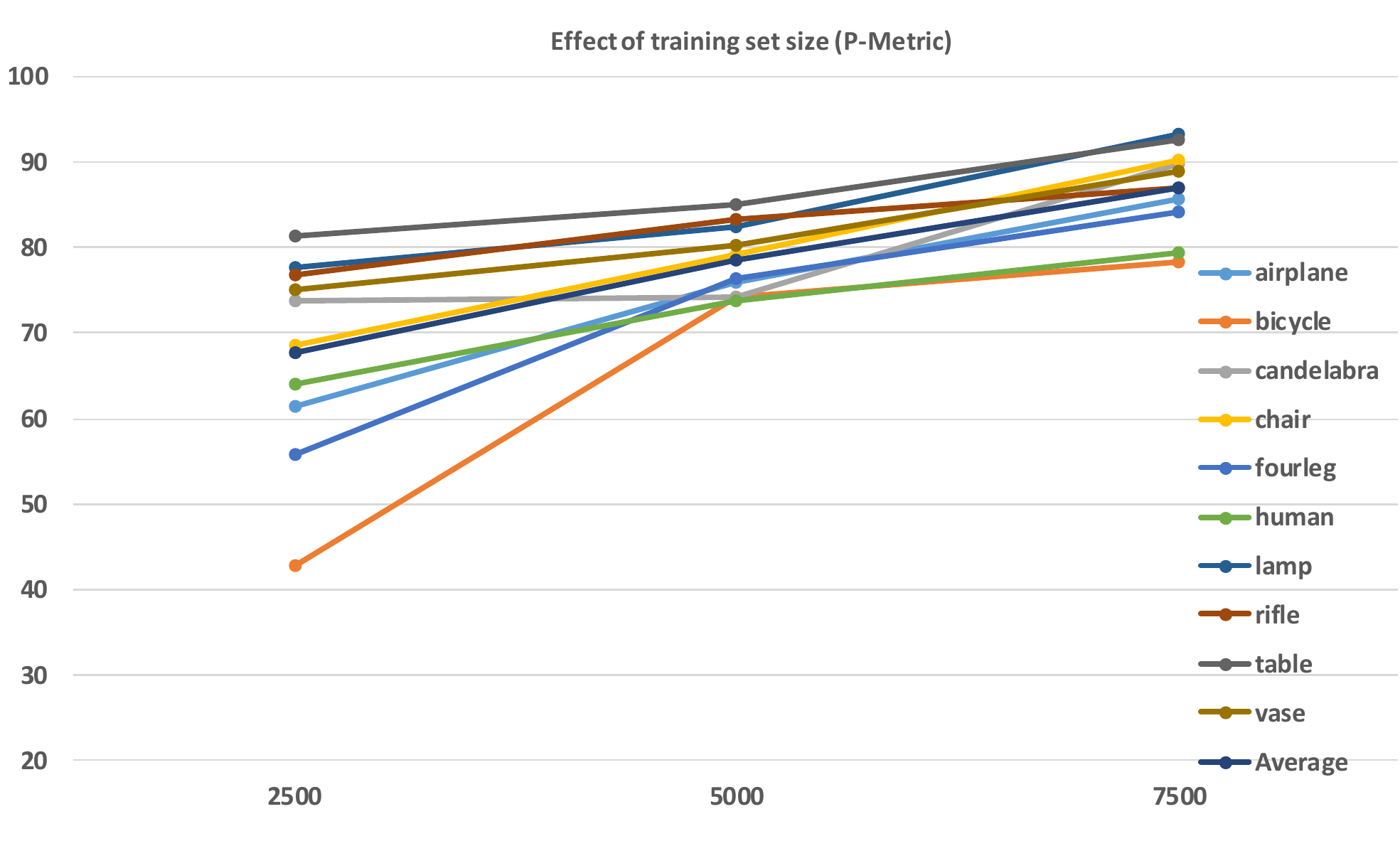}
	\end{subfigure}
	\begin{subfigure}{0.45\textwidth}
		\includegraphics[width = 1\textwidth]{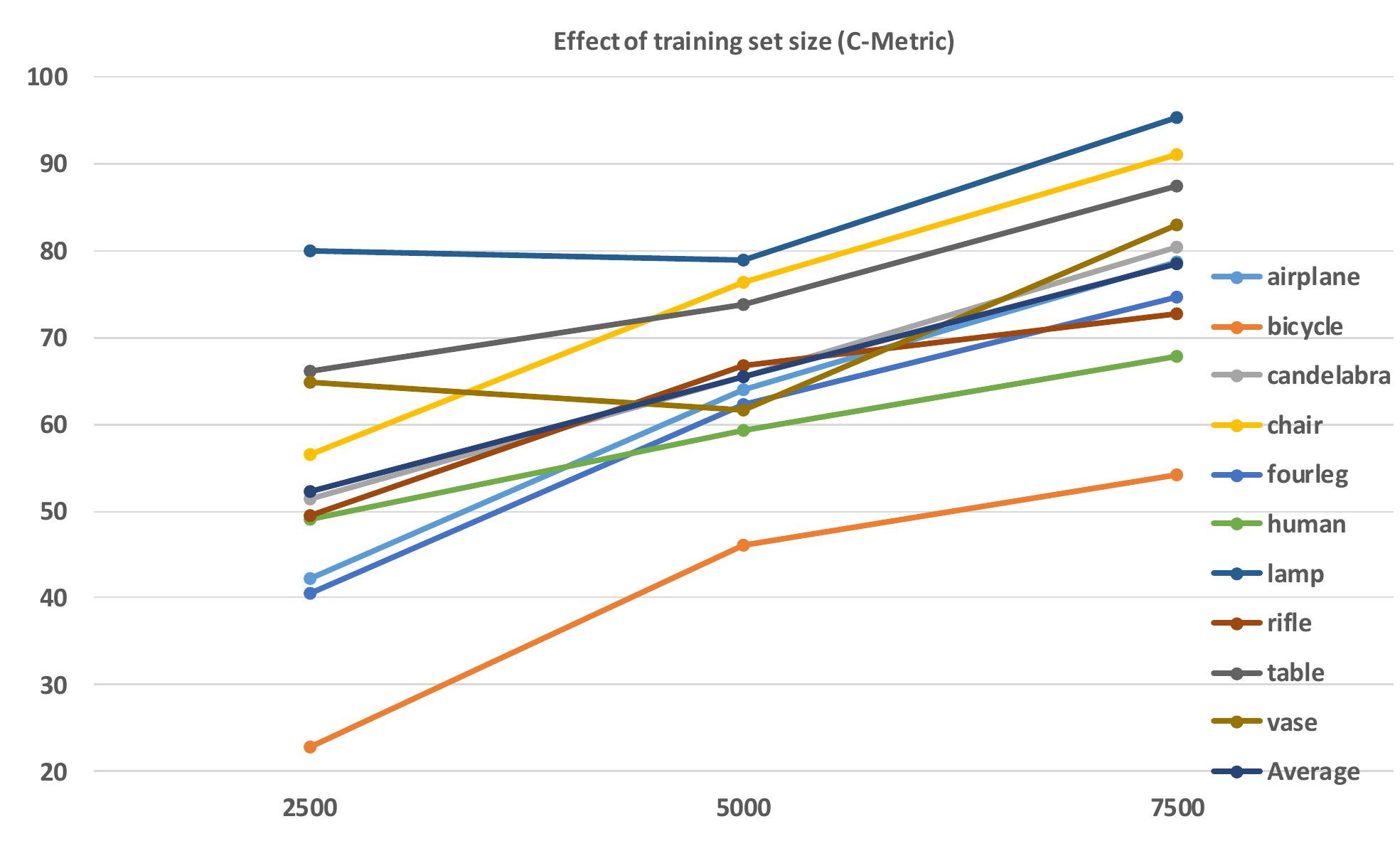}
	\end{subfigure}
	\caption{Segmentation performance of MCPNet with training sets of different sizes, evaluated by the P-metric (left) and C-metric (right).}
	\label{fig:trainSet}
\end{figure*}

To further investigate the robustness of the proposed MCPNet, we consider to change the ratio of training sketches in our experiments. Specifically, we have conducted experiments selecting 2,500, 5,000 and 7,500 sketches from each class as the training examples, and the results are presented in Figure \ref{fig:trainSet}. The overall performance of MCPNet is improved steadily as we increase the number of training sketches. The average segmentation accuracy of MCPNet evaluated by the P-metric is about $68\%$ when we use 1/4 sketches in our dataset for training, while that with 3/4 sketches is about $87\%$. Similar results can be observed using the C-metric. These experiments do indicate that the proposed MCPNet can generalize well and is a pragmatic solution for sketch segmentation.

\subsection{Ablation Study}

\begin{table*}[!ht]
\footnotesize
\newcolumntype{Y}{>{\centering\arraybackslash}X}
  \centering
  \caption{Accuracy (\%) of the proposed network with the different numbers of column.}\label{tab:diffscale}
\setlength{\tabcolsep}{1mm}{
  \begin{tabular}{p{1.5cm}|p{1.6cm}p{1.6cm}p{1.6cm}|p{1.6cm}p{1.6cm}p{1.6cm}}
        \toprule
        \hline
        &\multicolumn{3}{c|}{Pixel Metric}&\multicolumn{3}{c}{Component Metric}\\
        \cline{2-7}
        & {MCPNet-1}& {MCPNet-2} & {MCPNet-3}&{MCPNet-1}&{MCPNet-2}& {MCPNet-3}\\
        \hline
        airplane	&	82.8	&	77.4	&	{\bfseries85.8}	&	60.6	& 70.9 &	{\bfseries78.8} \\

        bicycle	&	72.4	&	75.2	&	{\bfseries78.3}	& 	44.3	&	{\bfseries59.9}	&	54.3 \\

        candelabra	&	86.2	&	86.0	&	{\bfseries89.8}	&	73.9	&	80.3	&	{\bfseries80.5} \\

        chair	&	87.1	&	88.8	&	{\bfseries90.2}	&	90.9	&	85.5	&	{\bfseries91.1}	\\

        fourleg	&	82.6	&	82.1	&	{\bfseries84.2}	&	72.3	&	{\bfseries76.8}	&	74.8 \\

        human	&	74.0	&	77.5	&	{\bfseries79.5}	&	60.4	&	66.5	&	{\bfseries67.8}	\\

        lamp	&	91.4	&	91.5	&	{\bfseries93.2}	&	92.5	&	91.2	&	{\bfseries95.5}	\\

        rifle	&	82.2	&	86.4	&	{\bfseries87.1} &	52.6	&	66.7	&	{\bfseries72.8}	\\

        table	&	87.5	&	88.1	&	{\bfseries92.6}	&	77.6	&	78.0	&	{\bfseries87.4}	\\

        vase	&	85.8	&	86.9	&	{\bfseries88.9}	&	78.3	&	79.2	&	{\bfseries83.1}	\\

        \hline
        Average	&	83.2	&	84.0	&	{\bfseries87.0}	&	70.8	&	74.5	&	{\bfseries78.6}	\\
          \hline
            \bottomrule
  \end{tabular}}
\end{table*}

One evident advantage of the proposed MCPNet is that it has adopted the multi-column strategy to extract effective features on different scale levels. To demonstrate this, we evaluate the performance of MCPNet with different numbers of columns, as shown in Table \ref{tab:diffscale}. It is easy to see that the segmentation accuracy is positively related to the number of columns, especially when we consider the C-metric: the average score of MCPNet-1 is $70.8\%$ while that of MCPNet-3 is $78.6\%$, and on the ``rifle" class we get the highest performance gain beyond $20\%$. Furthermore, the concatenation of inner features from the front end and back end works well, as MCPNet with only one column outperforms PointNet by averagely $3\%$ and $3.5\%$ with P-metric and C-metric, respectively.

\begin{table}[!ht]
\footnotesize
\newcolumntype{Y}{>{\centering\arraybackslash}X}
  \centering
  \caption{Time efficiency evaluated by the average processing time of one sketch (in ms).}\label{tab:avgtime}
\setlength{\tabcolsep}{1mm}{
  \begin{tabular}{p{1.35cm}|p{1.2cm}p{1.2cm}p{1.2cm}|p{1.2cm}p{1.45cm}p{1.45cm}p{1.45cm}}
        \toprule
        \hline

         Method & {U-Net}&{LinkNet}&{FCN}&{PointNet}&{MCPNet-1}&{MCPNet-2}&{MCPNet-3}\\
         \hline
         airplane & 125.6&165.2&168.4&94.8&{\bfseries74.4}&153.2& 236.8 \\
         bicycle &93.6&178&172.8&81.6&{\bfseries77.2}&122.4& 208.8 \\
         candelabra &129.6&205.6&153.2&{\bfseries81.2}&81.6&149.2&247.6 \\
         chair  & 160&169.6&164&85.2&{\bfseries70.4}&145.6& 190.4 \\
         fourleg  & 159.2&205.2&163.6&93.2&{\bfseries75.6}&162.8&192.8 \\
         human & 136&195.2&151.2&86&{\bfseries82.4}&159.2& 181.2 \\
         lamp & 129.6&188.8&144.4&84.8&{\bfseries74.8}&180.8&225.6 \\
         rifle & 126.4&212.8&164.4&82&{\bfseries77.6}&166& 224.4 \\
         table &130&172.4&159.2&86.8&{\bfseries77.6}&149.2& 203.2 \\
          vase & 133.6&190.8&138&86.8&{\bfseries83.6}&120.4& 182 \\
          \hline
          Average&132.36&188.36&157.92&86.24&{\bfseries77.52}&150.88&209.28\\
        \hline
    \bottomrule
  \end{tabular}}
\end{table}

Furthermore, taking point sets as input can reduce the computational cost and eventually make the multi-columns strategy become acceptable. For comparison, we calculate the average time of processing a sketch via different networks. From the results shown in Table \ref{tab:avgtime}, we can see that the computational cost of traditional DNNs designed for natural images is about twice that of PointNet and MCPNet-1. Nevertheless, their P-metric values are similar to that of PointNet. This validates our opinion that processing the blank space in sketch is unnecessary. Also, the computational cost of MCPNet-1 is slightly smaller than that of PointNet, that is because our MCPNet does not include the spatial transformation blocks in the PointNet. By comparing the results of MCPNets, we can find that the computational cost of the proposed network mainly depends on the number of columns, and adding one column will increase the processing time by about $70$ ms. To achieve the balance between efficiency and effectiveness, MCPNets with two or three columns are the considerable choice.

\subsection{Qualitative Analysis}

\begin{figure*}[!h]
  \centering
  \includegraphics[width=1\linewidth]{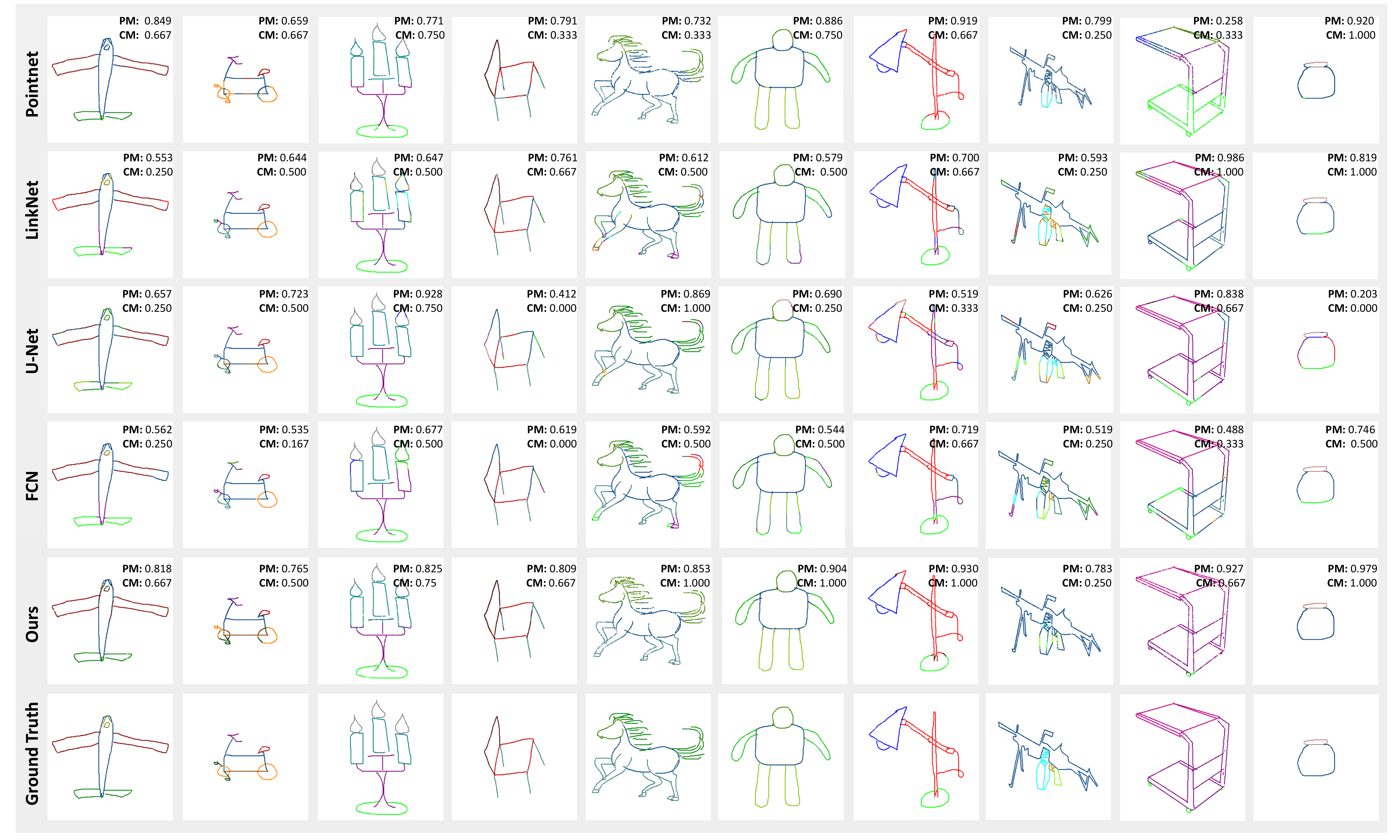}
  \caption{Qualitative examples of the proposed MCPNet. Each semantic component is annotated by a unique color, best viewed on a high-resolution screen.}\label{fig:examplesketch}
\end{figure*}

At the end of this section, we demonstrate some segmentation results for qualitative analysis. We select one example from each of the 10 classes on SketchSeg, and draw the results of baselines and MCPNet in Figure \ref{fig:examplesketch}. Note that in our experiments, sketches from all 10 classes are used together for training, thus it is unavoidable that a semantic component would be erroneously detected in sketch belonging to unrelated class. For example, FCN mistakenly considers the tail of a horse as the tube of a lamp, and the LinkNet considers part of the candle in a candelabra sketch as the magazine of a rifle. However, our MCPNet has fused features with various scales, which helps to alleviate such problem significantly. Our MCPNet even achieves a C-metric value of $1$ on the examples of horse, human and lamp.

\begin{figure}[h]
  \centering
  \includegraphics[width=1\linewidth]{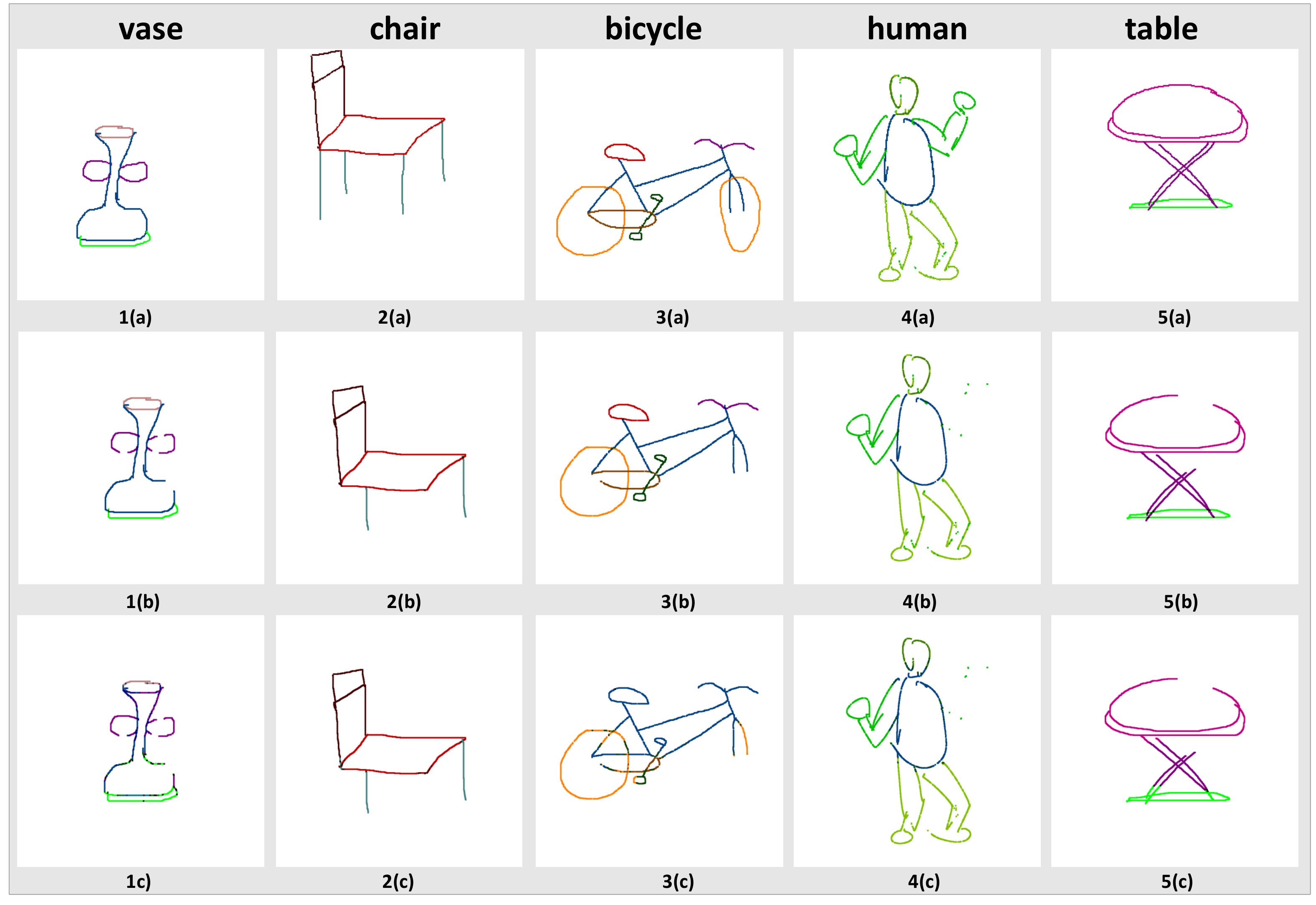}
  \caption{Exemplar segmentation results of noisy sketches. From top to bottom are original sketches, noisy sketches and segmentation results  generated by our method, respectively.}\label{fig:examplenoise}
\end{figure}

At last, considering that in real-life applications, sketches could be disrupted due to various factors, such as missing components, or abrupt, salt-and-pepper like noise, we further conduct an experiment on sketches with noise. We generate noisy sketches by the following two steps: we (i) manually remove one part of a sketch, and (ii) randomly add dots in the background, as shown in Figure \ref{fig:examplenoise}. Such synthetic noisy sketches are typical, yet the proposed MCPNet still handles these sketches well. This result suggests that MCPNet is robust to outliers, and it is a desirable solution for the sketch segmentation problem.

\section{Conclusion and Future Work}
We have introduced a new MCPNet for sketch segmentation, along with a new large SketchSeg dataset for training and evaluating deep learning-based methods. Our MCPNet is able to capture the sparse spatial structure information of sketches by learning and aggregating multi-scale deep representations on the sampled point sets, and extensive experimental results on the SketchSeg dataset have validated its effectiveness. In the future, we will further improve the SketchSeg dataset by introducing more categories and developing new evaluation metrics. We will also explore the new possible network architecture, e.g., encoding the order of strokes, to better tackle the sketch segmentation problem.

\bibliography{mybibfile}

\end{document}